\documentclass[11pt,preprintnumbers,aps,amssymb,nofootinbib,amsmath,superscriptaddress,notitlepage]
{revtex4-1}
\usepackage{epsfig,epsf}
\usepackage{bm} 
\usepackage{color} 
\usepackage{slashed}
\usepackage{relsize}	
\usepackage{soul} 
\usepackage{hyperref}
\usepackage{tensor} 
\newcommand{\beq}{\begin{equation}}
\newcommand{\beql}[1]{\begin{equation}\label{#1}}
\newcommand{\eeq}{\end{equation}}
\def\bal#1\gal{\begin{align}#1\end{align}}
\newcommand{\ball}[1]{\bal\label{#1}}
\newcommand{\eq}[1]{(\ref{#1})}
\newcommand{\fig}[1]{Fig.~\ref{#1}}
\renewcommand{\sec}[1]{Sec.~\ref{#1}}
\renewcommand{\b}[1]{{\bm #1}} 
\newcommand{\unit}[1]{\hat {{\bm #1}}} 

\newcommand{\aver}[1]{\left\langle #1 \right\rangle}

\begin{document}

\title{Anomalous scattering and transport  in chiral matter}

\author{Kirill Tuchin}

\affiliation{
Department of Physics and Astronomy, Iowa State University, Ames, Iowa, 50011, USA}

\date{\today}


\begin{abstract}
 
Chiral anomaly modifies the scattering processes in chiral systems. These processes can be investigated using the Maxwell-Chern-Simons theory that couples electrodynamics to the pseudoscalar field $\theta$ describing the topological charge induced by external sources. Assuming slow variation of the topological charge density,  the fermion scattering cross section is computed in the Born approximation and is found to have a resonance at the scattering angles proportional to the chiral conductivity. As a result, the transport coefficients are suppressed at high temperatures, which may be a reason for the smallness of the  Quark Gluon Plasma viscosity. The anisotropy of the cross section arises due to the spatial  variation of the topological charge; its effect on the electrical conductivity is discussed.

\end{abstract}

\maketitle

\section{Introduction}\label{sec:a}

Systems containing chiral fermions are very diverse and posses a number of intriguing properties that are  revealed when the topological charge is induced in them \cite{Zhitnitsky:2012ej,Kharzeev:2007tn,Kharzeev:2015znc,Li:2014bha,Gorbar:2018nmg,Marsh:2015xka,Klinkhamer:2004hg}. In particular, in the presence of the topological charge the Green's functions of the electromagnetic field acquire $P$ and $CP$-odd components \cite{Carroll:1989vb} which manifest as novel fermion interactions.

This paper considers scattering of electrically charged particles in chiral systems with finite topological charge. The scattering is mediated by electromagnetic field excitations some of which are $P$ and $CP$ even and others are odd. The latter induce novel terms in the scattering cross section that can be studied experimentally. They also affect the transport cross sections and thus the kinetic coefficients of chiral systems which are also of certain practical interest. The results of this paper pertain to any chiral system, however, for the sake of semantic simplicity it is convenient to refer to  light charged fermions as ``electrons" and to heavy charged fermions as ``ions". 

The new scattering effects in chiral matter have the most clear interpretation when ``ions" are non-relativistic.  More precisely, we consider scattering of ``electrons" on heavy ``ions" assuming that the momentum transfer is much smaller than the ion mass $M$, which will be referred to as the static limit.  It is actually not difficult to obtain fully relativistic results, however they are quite bulky and hardly illuminating. 

In the Born approximation, the scattering amplitude is proportional to the potential energy in the momentum space. The latter is encoded in the photon propagator which can be derived using the Maxwell-Chern-Simons theory \cite{Wilczek:1987mv,Sikivie:1984yz} that couples electrodynamics to the topological charge induced by external sources. The corresponding term in the Lagrangian is 
\cite{Wilczek:1987mv,Carroll:1989vb, Sikivie:1984yz}
\ball{i1}
\mathcal{L}_A =-\frac{c_A}{4}\theta F_{\mu\nu}\tilde F^{\mu\nu}\,,
\gal
where $c_A$ is the  chiral anomaly coefficient of QED  \cite{Adler:1969gk,Bell:1969ts} and the dimensionless pseudoscalar field $\theta$ describes the topological charge. In many systems $\theta$ is believed to be a slowly varying function of coordinates and time \cite{Kharzeev:2015znc}. This is the approximation also assumed in this work. In particular, we treat the first derivative $\partial\theta$ as constant and adopt a fairly standard notation of the constant vector $b^\mu= (b_0,-\b b)=c_A\partial^\mu \theta= c_A(\dot \theta, -\b \nabla \theta)$; $b_0$ is also known as the chiral conductivity $\sigma_\chi$  \cite{Kharzeev:2009pj,Fukushima:2008xe}.
The photon propagator in chiral matter reads \cite{Carroll:1989vb,Lehnert:2004hq}
\ball{b10}
D_{\mu\nu}(q)= -i \frac{q^2 g_{\mu\nu}+i\epsilon_{\mu\nu\rho \sigma}b^\rho q^\sigma+b_\mu b_\nu}{q^4+b^2 q^2-(b\cdot q)^2}\,.
\gal
 Since we are interested in the static limit, it is convenient to introduce a notation $D_{\mu\nu}(\b q)= \lim_{q^0\to 0} D_{\mu\nu}(q)$ and the corresponding expression in the configuration space $D_{\mu\nu}(\b x)$. Throughout the paper the bold face font distinguishes the three-dimensional vectors. The potential induced by a stationary current $J^\nu(\b x)$ can be computed as  
\ball{b11}
A^\mu (\b x) = -i\int d^3x' D^{\mu\nu}(\b x-\b x')J_\nu(\b x')= -i\int \frac{d^3q}{(2\pi)^3} e^{i\b q\cdot \b x}D^{\mu\nu}(\b q) J_\nu (\b q)\,.
\gal

In the forthcoming sections we consider electron scattering off potential $A^\mu$ at the leading order of the perturbation theory.  Following \cite{Qiu:2016hzd} it is instructive to consider two different types of chiral matter: (i) homogenous matter with $b_0\neq 0$, $\b b=0$ and (ii) stationary matter with $b_0=0$, $\b b\neq 0$. The paper is organized accordingly: \sec{sec:b} deals with the homogeneous chiral matter in which case the scattering cross section \eq{d5} is found to have a resonance at momentum transfer $q^2=- b_0^2$. It appears due to the periodic variation of the vector potential with the wavenumber $b_0$\cite{Qiu:2016hzd,Tuchin:2014iua} and is intimately related to the chiral instability of the electromagnetic field \cite{Carroll:1989vb}. Therefore, at $T\gg b_0$ the transport cross section $\sigma_T$ is enhanced as can be seen in \eq{d8}. As a consequence, the mean free path, which in a dilute gas of density $n$ can be estimated as  $\ell \sim 1/n\sigma_T$, is shorter than at $b_0=0$ by a factor of $\sim M^2/T^2$. This implies suppression of transport coefficients at high temperatures and, in particular, of the ratio of the shear viscosity to the entropy density is $\eta/s\sim T\ell \aver{v}$, which may be a reason for the smallness of this ratio in Quark Gluon Plasma \cite{Heinz:2013th}. Indeed, the gluon propagator has exactly same form as \eq{b10} apart from the color factor. 

The stationary chiral matter is discussed in \sec{sec:d} and the corresponding scattering and transport cross sections given by \eq{c8} and \eq{c10} reflect the axial symmetry with respect of the vector $\b b$.  In \sec{sec:j} the effect of the new terms in the transport cross section is illustrated by computing the electrical conductivity  using the classical transport theory. In homogeneous matter the result is displayed in \fig{fig9} which shows suppression of the conductivity at $T\gg b_0$. In stationary matter the applied external electric field induces electric current along its direction and in the direction of $\b b$ with the corresponding conductivities $\sigma$ and $\sigma'$. Their temperature dependence is shown in \fig{fig5}. The discussion and conclusions are presented in \sec{sec:s}.

\section{Homogenous matter}\label{sec:b}

\subsection{Potential}

In homogeneous chiral matter with $\b b=0$, $b_0\neq 0$, the components of the propagator \eq{b10} read in the static limit \cite{Qiu:2016hzd}
\begin{subequations}\label{b12}
\bal
&D_{00}(\b q)= \frac{i}{\b q^2}\,,\label{b12a}\\
&D_{0i}(\b q)= D_{0i}(\b q)= 0\,,\label{b12b}\\
&D_{ij}(\b q)=-\frac{i\delta_{ij}}{\b q^2-b_0^2}-\frac{\epsilon_{ijk}q^k}{b_0(\b q^2-b_0^2)}+\frac{\epsilon_{ijk}q^k}{b_0\b q^2}\,.\label{b12c}
\gal
\end{subequations}
The current density of the static point source (the ``ion") of charge $e'$ is $J^\nu(\b x) =e' \delta\indices{^\nu_0}\delta(\b x)$. It induces the Coulomb potential 
\ball{b15}
A^0(\b q)= e'/\b q^2\,,\qquad \b A(\b q)=0
\gal
implying that the scattering cross section off the point charge is given by the Rutherford formula and is not affected by the anomaly (in the static limit).  

A non-trivial contribution comes about if the ``ion" is in a state $\psi$ with a finite expectation value of the magnetic moment $\b \mu$. Indeed, the spin current associated with such a state is $\b\nabla \times \psi^* \b \mu \psi$. In the point particle limit the spin current can be written as  $\b J(\b x) = \b \nabla \times (\b \mu\delta(\b x))$. It represents the first non-vanishing  multipole moment of the vector potential. Altogether the electrical current of ion is
\ball{b16}
J^0(\b x) =e'\delta(\b x)\,,\qquad 
\b J(\b x) = \b \nabla \times (\b \mu\delta(\b x))\,,
\gal
which in momentum space reads
\ball{b17}
J^0(\b q)=e'\,,\qquad \b J(\b q)= i\b q\times \b \mu\,.
\gal
According to \eq{b11} and  \eq{b12} it  produces the potential 
\begin{subequations}\label{b19}
\bal
A^\ell(\b q)&=-iD^{\ell i}(\b q)J_i(\b q)=  -\epsilon_{ijk}\mu^kq^j\left( -\frac{i\delta_{\ell i}}{\b q^2-b_0^2}-\frac{\epsilon_{\ell i r}q^r}{b_0(\b q^2-b_0^2)}+\frac{\epsilon_{\ell i r}q^r}{b_0\b q^2}\right)\label{b19a}\\
&
= -\frac{1}{\b q^2-b_0^2}\left[ i(\b \mu\times \b q)^\ell + \frac{b_0}{\b q^2}(\b \mu\cdot \b q q^\ell-\b q^2 \mu^\ell)\right]\,, \label{b19b}
\gal
\end{subequations}
while the time component is still given by the first equation of \eq{b15}. Transformation to the configuration space is accomplished using the integral
\ball{b20}
\int \frac{e^{i\b q\cdot \b x}}{|\b q|^2-b_0^2-i0}d^3q=
 \frac{2\pi^2}{|\b x|}e^{i|\b x|b_0}\,.
\gal
The potential,   or, more precisely, the zero-frequency component of the vector potential, in the configuration space reads
\ball{b22}
\b A(\b x)=&\frac{\b\mu \times \b x}{4\pi |\b x|^3}\left( 1-ib_0|\b x|\right)e^{ib_0|\b x|}-\frac{\b\mu}{2\pi b_0|\b x|^3}\left( 1-e^{ib_0|\b x|}+ib_0|\b x|e^{ib_0|\b x|}\right)\nonumber\\
&+\frac{(\b\mu\times \b x)\times \b x}{4\pi b_0 |\b x|^5}\left[3\left(e^{ib_0|\b x|}-1\right) -ib_0|\b x|(3-ib_0|\b x|)e^{ib_0|\b x|}\right]\,.
\gal
 In the anomaly-free limit  $b_0\to 0$ \eq{b22} reduces to the classical result
\ball{b24} 
\b A(\b x)= \frac{1}{4\pi}\frac{\b \mu\times\b x}{|\b x|^3}= \frac{1}{4\pi}\int \frac{\b \nabla'\times (\b \mu\delta(\b x'))}{|\b x-\b x'|}d^3x'\,, \qquad b^\mu=0\,.
\gal
Since $\mu\sim e/M$, the magnetic contribution \eq{b22} is a relativistic correction to the Coulomb potential. 

 The oscillatory behavior of the potential \eq{b22} stems from the non-dissipative nature of the anomalous current \cite{Kharzeev:2011ds}. Its imaginary part indicates the radiative instability  with respect to the pair-production \cite{Tuchin:2018sqe}. An important feature of the photon propagator  in chiral matter \eq{b10} is the emergence of the unstable modes that produce the chiral instability of the electromagnetic field. This instability originates in the momentum interval $|\b q|<b_0$ and appears as the pole in the upper-half of the complex $q^0$-plane \cite{Carroll:1989vb,Tuchin:2014iua}. In the static limit $q^0\to 0$ there is a single mode $|\b q|=b_0$ that causes the chiral instability and it appears as the singularity in \eq{b19b}.

\subsection{Cross sections}

The scattering cross section of ``electron" of charge $e$ off an ``ion" of charge $e'$ reads
\ball{c3}
\frac{d\sigma}{d\Omega'} = \frac{e^2}{8\pi^2}\left[ p\cdot A^*(\b q) \, p'\cdot A(\b q)-p\cdot p' |A(\b q)|^2+p\cdot A(\b q)\, p'\cdot A^*(\b q)+m^2|A(\b q)|^2\right]\,,
\gal
where $p=(E,\b p)$ and $p'=(E,\b p')$ are particle momentum before and after scattering and $\b q=\b p'-\b p$ is the momentum transfer. In terms of the scattering angle $\vartheta$ the momentum transfer is $|\b q|= 2|\b p|\sin\frac{\vartheta}{2}$. To be sure, the electron wave function also gets anomalous contributions, however those external leg contributions are irrelevant for the scattering problem. 

 The static limit which is used in derivation of \eq{c3} requires that the energy transfer be negligible compared to (i) the momentum transfer, i.e.\ $q^0\ll |\b q|$ and (ii) the anomaly parameter, i.e.\  $q^0\ll b_0$ as can be seen by examining the denominator of \eq{b10}. Using  $q_0= \b q^2/2M$ the first condition  implies  that $|\b q|\ll M$, while  the second condition imposes a stronger constraint $|\b q|\ll \sqrt{b_0M}$ (since $b_0\ll M$). In the static limit the interaction time $1/q_0$ is much longer than the time $1/b_0$ it takes the chiral instability to develop.

Substituting the vector potential from\eq{b19b} and the scalar potential from \eq{b15} into \eq{c3} yields
\ball{d1}
\frac{d\sigma}{d\Omega'}= \frac{e^2}{4\pi^2}
\left\{
\left[ \frac{e'E}{\b q^2}+\frac{b_0(\b \mu\cdot \b q\, \b p\cdot \b q- \b p\cdot \b \mu \, \b q^2)}{(\b q^2-b_0^2)\b q^2}\right]^2 + \frac{[\b \mu\cdot (\b p\times \b p')]^2}{(\b q^2-b_0^2)^2}
-\frac{\b q^2}{4}\left[ \frac{e'^2}{\b q^4}-\frac{(\b \mu\times \b q)^2}{(\b q^2-b_0^2)^2}\left(1+\frac{b_0^2}{\b q^2}\right)\right]
\right\}\,.
\gal
Since the magnetic moments are usually randomly oriented,  Eq.~\eq{d1} needs to be averaged over its directions. Using $\aver{\mu_i\mu_j}= \mu^2\delta_{ij}/3$ one finds
\ball{d5}
\aver{\frac{d\sigma}{d\Omega'}}= \frac{e^2}{8\pi^2}\left\{
\frac{2E^2e'^2}{\b q^4}\left( 1-\frac{\b q^2}{4E^2}\right)+\frac{2\mu^2}{3(\b q^2-b_0^2)^2}\left(1+\frac{b_0^2}{\b q^2}\right) \left[ (\b p\times \b q)^2+\frac{\b q^4}{2}\right]
\right\}\,.
\gal
The first term in \eq{d5} corresponds to scattering off the Coulomb potential, while the second one involves a contribution of the magnetic moment. It is the last contribution that is dependent on the anomaly parameter $b_0$. It produces the resonant behavior at momentum transfers $\b q^2= b_0^2$. The origin of this behavior can be seen in the coordinate space, where the vector potential \eq{b22} oscillates with the wave number $b_0$. As have been already noted, $|\b q|= b_0$ is the runaway mode responsible for the chiral instability.

The transport cross section is defined as 
\ball{c4}
\sigma_T= \int (1-\cos\vartheta)d\sigma = \frac{1}{2\b p^2}\int \b q^2 d\sigma \,.
\gal
Integration on the right-hand-side runs over the directions of momentum $\b p'$.
Plugging \eq{d5} into \eq{c4} and using $(\b p\times \b q)^2 = \b p^2\b q^2-\b q^4/4$ we obtain for the transport cross section
\ball{d7}
\sigma_T= 
 \frac{e^2}{16\pi \b p^4}\int_0^{(2p)^2}\left\{
\frac{2E^2e'^2}{\b q^2}\left( 1-\frac{\b q^2}{4E^2}\right)+\frac{2\mu^2}{3}\frac{\b q^2}{(\b q^2-b_0^2)^2+\Gamma^4}\left(1+\frac{b_0^2}{\b q^2}\right) \left( \b p^2 \b q^2+\frac{\b q^4}{4}\right)
\right\}d\b q^2
\gal
where $\Gamma$   is related to the photon decay rate $w$ in the chiral medium as $\Gamma^2= q_0w$  where $w\sim e^2b_0$ \cite{Tuchin:2018sqe}. Actually, any process that tames the chiral instability also contributes to $w$. 

\begin{figure}[ht]
      \includegraphics[height=5cm]{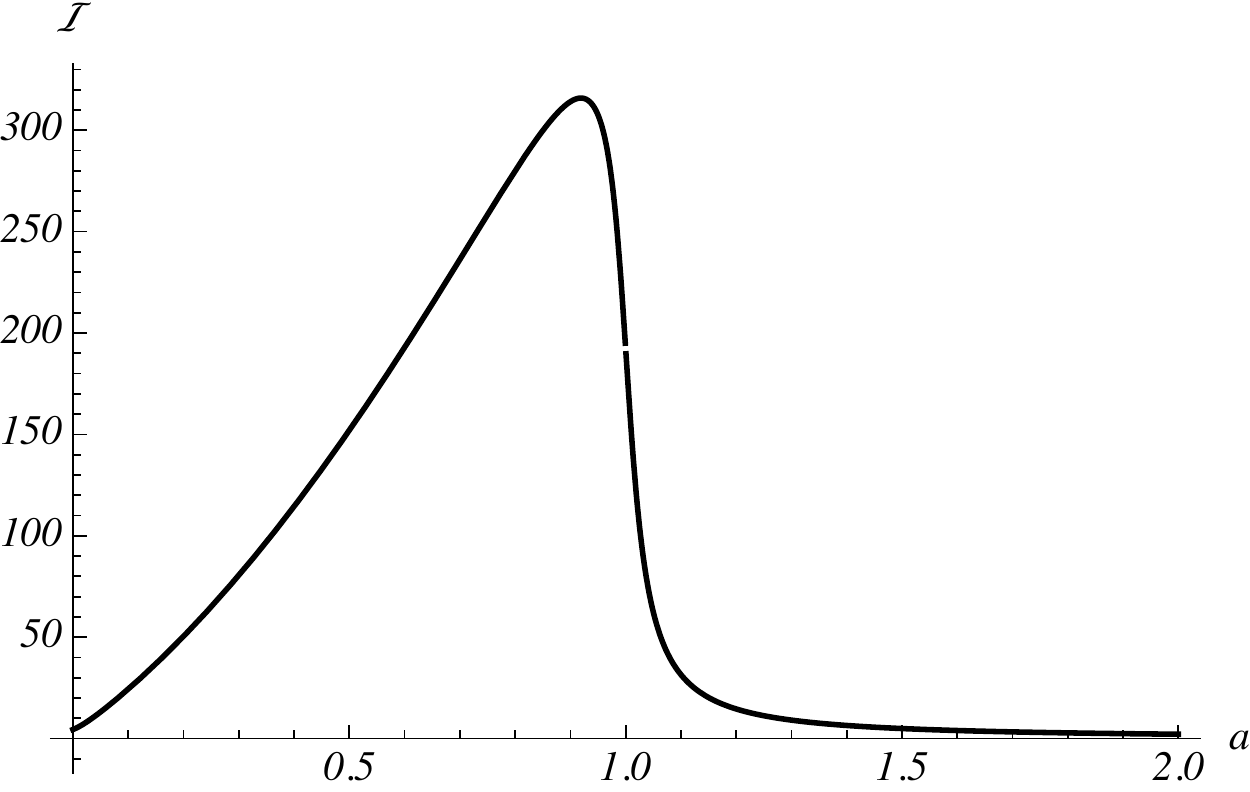} 
  \caption{Function $\mathcal{I}(a)$ at $\epsilon=0.03$.}
\label{fig3}
\end{figure}

Integration in \eq{d7} yields 
\ball{d8}
\sigma_T=\frac{e^2}{16\pi \b p^4}\left( 4E^2e'^2L+ \frac{2\mu^2}{3}4\b p^4\mathcal{I}\right)\,,
\gal
where $L$ stands for the Landau logarithm  (the first, Coulomb term, in \eq{d7} is computed with the logarithmic accuracy) and we defined 
\ball{d9}
\mathcal{I}= 1+\frac{1}{\epsilon}\left[ 2a(1+a)-\epsilon^2\right]\left( \arctan\frac{1-a}{\epsilon}+\arctan\frac{a}{\epsilon}\right)+\frac{1}{2}(1+3a)\ln \left(1+\frac{1-2a}{a^2+\epsilon^2}\right)\,,
\gal
with 
\ball{d10}
a= \frac{b_0^2}{4\b p^2}\,,\qquad \epsilon= \frac{\Gamma^2}{4\b p^2}\,.
\gal
This function is displayed in \fig{fig3}. 
At large momenta  \eq{d9} can be expanded at small $a$ while keeping the ratio $a/\epsilon$ fixed. With the logarithmic accuracy this gives $\mathcal{I}\approx \ln a^{-1}$. If $|\b p|\gg M$, then the anomalous term dominates and the transport cross section which reads\footnote{A more precise expression is obtained by keeping sub-logarithmic corrections in \eq{d9} which is accounted for by replacing in \eq{d12} $\ln(4\b p^2/b_0^2)\to \ln(4\b p^2/b_0^2)+ 1+\pi y+2y \arctan y-(1/2)\ln(1+1/y^2)$ with $y=b_0^2/\Gamma^2$. \label{foot}}
\ball{d12}
\sigma_T\approx \frac{e^2\mu^2}{6\pi } \ln\frac{4\b p^2}{b_0^2} \,,\qquad |\b p|\gg b_0\,, \Gamma\,.
\gal
In the opposite limit $|\b p|\ll b_0,\Gamma$, $\mathcal{I}\approx 6\b p^2b_0^2/(b_0^4+\Gamma^4)$ so that the Coulomb term dominates the transport cross section. It is noteworthy that the transport cross section is only logarithmically sensitive to the chiral conductivity $b_0$ and does not depend on the external parameter $\Gamma$ in two asymptotic regimes. The transition between these regimes at moderate momenta is the most sensitive to their values.

\section{Stationary matter}\label{sec:d}

In stationary matter with $\b b\neq 0$, $b_0= 0$, the propagator components are \cite{Qiu:2016hzd}
\begin{subequations}\label{b33}
\bal
&D_{00}(\b q)=\frac{i\b q^2}{\b q^4+\b b^2\b q^2-(\b b\cdot \b q)^2}\,,\label{b34}\\
&D_{0i}(\b q)=-D_{i0}(\b q)= \frac{\epsilon_{ijk}b^jq^k}{\b q^4+\b b^2\b q^2-(\b b\cdot \b q)^2}\,,\label{b35}\\
&D_{ij}(\b q)=-i\frac{\b q^2\delta_{ij}+b_ib_j}{\b q^4+\b b^2\b q^2-(\b b\cdot \b q)^2}\,.\label{b36}
\gal
\end{subequations}
The anomalous contribution arises already from the leading multipole moment sourced by $J^0=e'\delta(\b x)$. The corresponding  potential $ A^\mu (\b q)= -ie'D^{\mu 0}(\b q)$ has components
\ball{b40}
A^0(\b q) = \frac{e'\b q^2}{\b q^4+\b b^2\b q^2-(\b b\cdot \b q)^2}\,,\qquad
\b A(\b q) = \frac{ie'\b b\times \b q}{\b q^4+\b b^2\b q^2-(\b b\cdot \b q)^2}\,.
\gal 
The expressions in the configuration space can be found in \cite{Qiu:2016hzd}.  This configuration is axially symmetric with respect to the vector $\b b$.

Employing \eq{b40} in \eq{c3} one derives the cross section
\ball{c8}
\frac{d\sigma}{d\Omega'} = \frac{(ee')^2}{16\pi^2}\frac{4E^2\b q^4 +4[\b p\cdot (\b b\times \b p')]^2-\b q^2[\b q^4-(\b b\times \b q)^2]}{ \b q^4(\b q+\unit q\times \b b)^4}\,,
\gal
where $\unit q= \b q/|\b q|$. Two features are the most noteworthy in \eq{c8} as compared to the Rutherford formula:  breaking of the axial symmetry by $\b b$ and disappearance of the small scattering angle divergence. The latter, explicitly seen  in \eq{c9},  is particularly significant in the kinetic theory.  We note the lack of the resonant  behavior, which is closely related to the absence of the chiral instability at $b_0=0$.

\begin{figure}[ht]
\begin{tabular}{cc}
      \includegraphics[height=5cm]{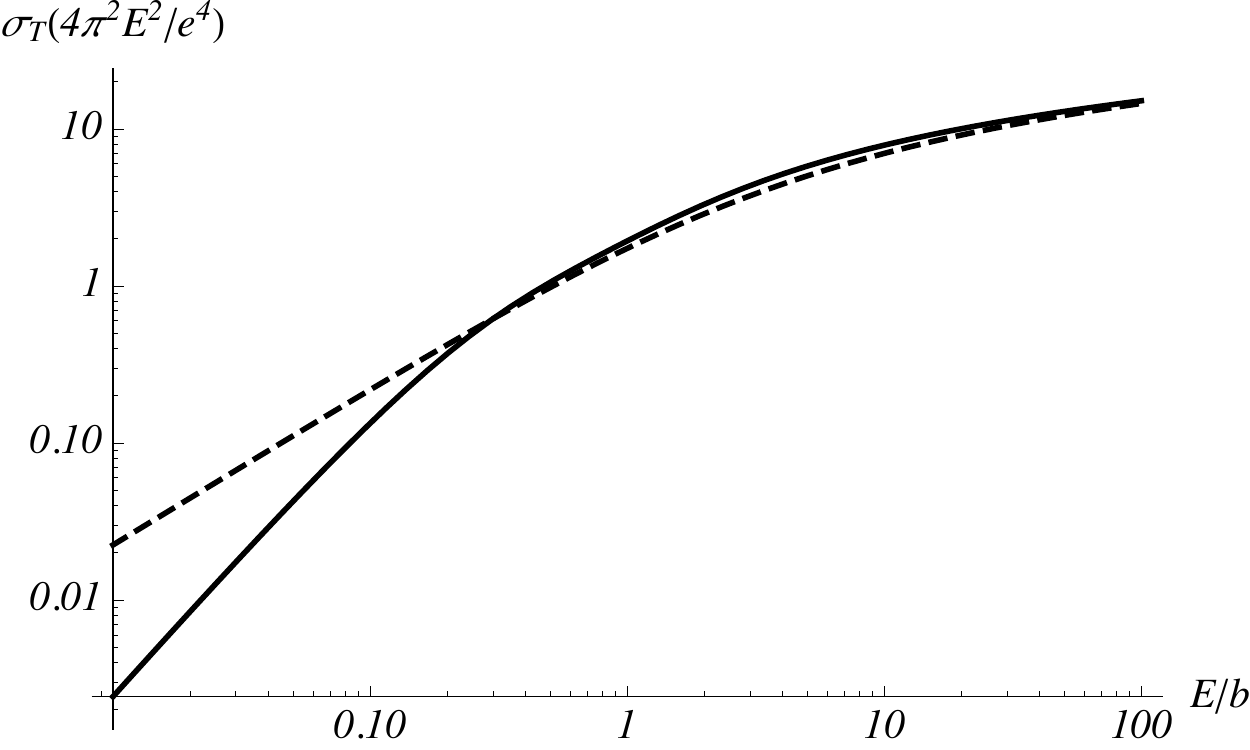} &
      \includegraphics[height=5cm]{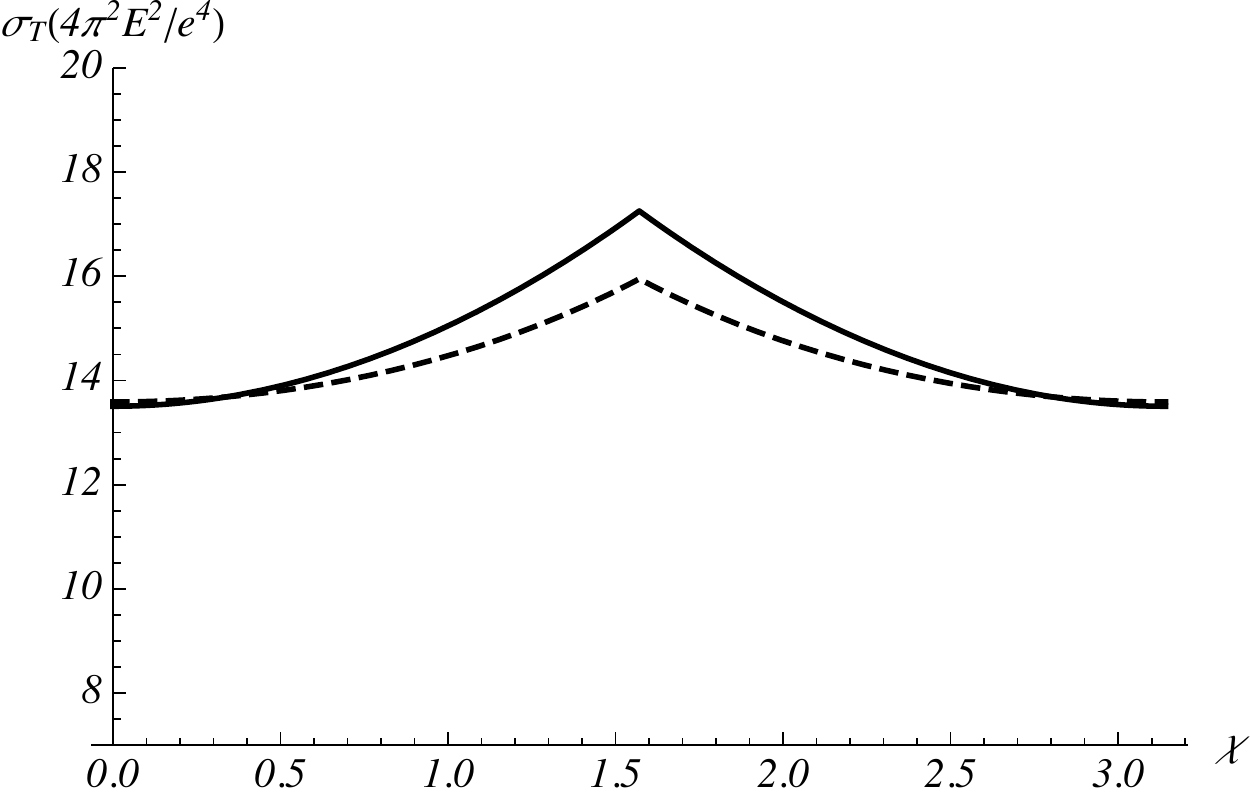}
      \end{tabular}
  \caption{The transport cross section at $\b b\neq 0$ and $b_0=0$ versus $E/b$ at $\chi=1$ (left panel) and versus $\chi$ at  $E/b=100$ (right panel). Solid line: exact result, dashed line: approximation \eq{c10}. }
\label{fig1}
\end{figure}

The transport cross section is obtained by substitution of \eq{c8} into \eq{c4}.  It is convenient to set up a Cartesian reference frame with the $z$-axis along the momentum $\b p$ and vector $\b b$ in the $yz$-plane so that the the coordinates of all relevant vectors can be parameterized as 
\ball{c6}
\b p = |\b p|(0,0,1)\,,\qquad \b p'= |\b p|(\sin\vartheta\cos\phi, \sin\vartheta\sin\phi, \cos\vartheta)\,,\qquad \b b= |\b b|(0,\sin\chi, \cos\chi)\,
\gal
and the solid angle $d\Omega' = d\cos\vartheta d\phi$.  
The angular integration  can be done analytically in an important case $|\b p|\gg |\b b|$ when the cross section is dominated by the small angle scattering. We have 
\begin{subequations}
\bal
\sigma_T\approx &\frac{(ee')^2}{16\pi^2}\int_0^{2\pi}d\phi  \int_0^{\sim 1}d\vartheta^2\frac{E^2\vartheta^2+b^2\sin^2\chi\cos^2\phi}{(p^2\vartheta^2+b^2-b^2\sin^2\chi\sin^2\phi)^2} \label{c9}\\
\approx & \frac{(ee')^2}{16\pi^2}\frac{E^2}{\b p^4}\int_0^{2\pi}\ln \frac{\b p^2}{\b b^2(1-\sin^2\chi\sin^2\phi)}d\phi=  \frac{(ee')^2}{4\pi}\frac{E^2}{\b p^4}\ln \frac{\b p^2}{\b b^2(1+|\cos\chi|)}\,.\label{c10}
\gal
\end{subequations}
In the limit $\b b\to 0$ the denominator in the argument of the logarithm is replaced by the Debye mass $m_D$ squared. \fig{fig1} shows the transport cross section calculated with \eq{c8} and an approximation \eq{c10}.

\section{Electrical conductivity}\label{sec:j}

The resonance in the scattering cross section at $|\b q|\approx  b_0\neq 0$ enhances the transport cross section \eq{d8} at $|\b p|\gtrsim b_0$ as seen in \fig{fig3}. Since the mean free path is inversely proportional to the transport cross section, at least in the dilute gas limit, the transport coefficients get suppressed at $T \gtrsim b_0$. In this section we employ the classical transport theory to illustrate this and other effects of the chiral anomaly on the electrical conductivity. We neglect the quantum statistical effects so that the equilibrium distribution of particles of energy $E$ at temperature $T= \beta^{-1}$  is proportional to $e^{-\beta E}$. The terms in the kinetic equations that involve the Berry curvature \cite{Son:2012zy} are likewise neglected as they are not essential for the purpose of this illustration. The main goal here is to estimate the effect of the anomaly on the collision integral. 

In the linear response approximation the electric current induced by constant external electric field $\bm{\mathcal{E}}$ is given by
\ball{k1}
\b j = -e\int \b v \delta f d^3p = \frac{e^2}{T}\int f_0\nu^{-1} (\bm{\mathcal{E}}\cdot \b v) \b v d^3p\,,
\gal
where $\nu = n' v_\text{rel}\sigma_T$ is the collision rate of electrons with ions of density $n'$, $v_\text{rel}$ is the relative velocity and $\b v = \partial E/\partial \b p$.  In the chiral limit, the equilibrium distribution of electrons $f_0$ is the one of free massless fermions because of the peculiar form of their dispersion relation $(E\pm b_0)^2=(\b p\mp \b b)^2$ (see Appendix). Calculations in this section are performed in the chiral limit and furthermore it is assumed that $e=e'$ and $n=n'$ for notational simplicity. We proceed by computing the current \eq{k1} in two limiting cases.

\subsection{Homogeneous matter}

\begin{figure}[ht]
      \includegraphics[height=5cm]{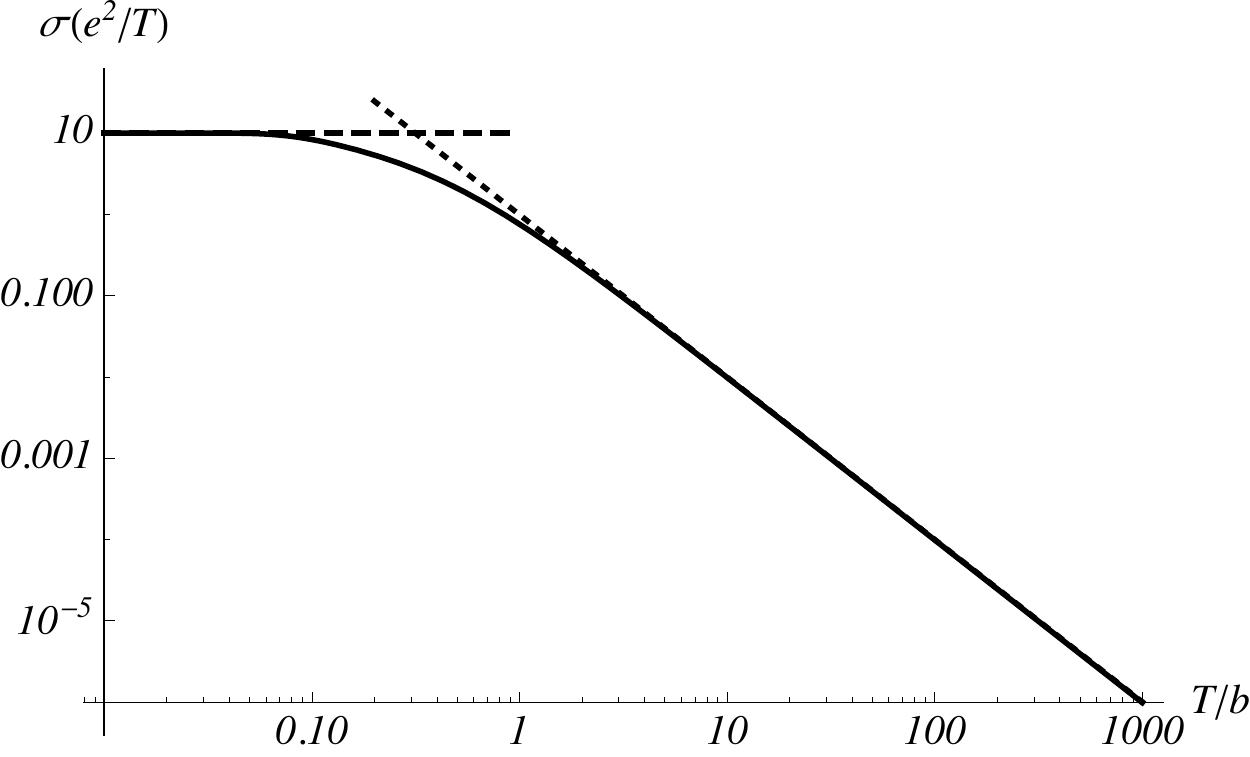} 
  \caption{Electrical conductivity \eq{k21} at $\Gamma= 0.1b_0$, $M= 5b_0$ and $L=5$  (solid line). Dashed line: the Coulomb limit, dotted line: the anomalous contribution \eq{k23} with the correction mentioned in footnote \ref{foot}. In relativistic heavy-ion collisions  $b_0$ is conjectured to be on the order of 1--10 MeV \cite{Kharzeev:2015znc}. }
\label{fig9}
\end{figure}

Electrical conductivity  at $\b b=0$ can be computed using the standard formula
\ball{k21}
\sigma = \frac{e^2}{3T}\int f_0 \frac{1}{n \sigma_T}d^3p \,,
\gal
where the transport cross section is given by \eq{d8}. As explained there, at  low energies/temperatures the Coulomb term dominates the transport cross section so that one recovers the textbook result $\sigma= 16\pi T/e^2 L$, where $L=\ln(T/m_D)$. At high temperatures one obtains using \eq{d12} 
\ball{k23} 
\sigma = \frac{\pi}{\mu^2 T \ln(T/b_0)}\,,\qquad T\gg b_0,\,\Gamma\,.
\gal
Remarkably,  this formula does not depend on the resonance width $\Gamma$ (which is essentially a cutoff) and only weakly depends on $b_0$. It shows that the electrical conductivity of homogenous chiral matter with $b_0\neq 0$ is suppressed at high temperatures by a factor of  $e^2/8\mu^2T^2\sim M^2/2T^2$, where $M$ is the ion  mass. This is exhibited in \fig{fig9}, neglecting the temperature dependance of the Landau logarithm. The suppression happens because of the resonance in the scattering cross section at the scattering angle $\vartheta= b_0/|\b p|$, which in turn can be traced back to the oscillatory behavior of the potential as explained in \sec{sec:b}. 

\subsection{Stationary matter}

At finite $\b b$ (and $b_0=0$) the current is no longer isotropic. It follows from \eq{k1} that the most general relationship between the current and the electric field has form
\ball{k3}
j_i=\Sigma_{ij}\mathcal{E}_j\,,
\gal
where $\Sigma_{ij}$ is a symmetric tensor. Since the only available symmetric tensor are $\delta_{ij}$ and $\hat b^i\hat b^j$  it can be expanded as
\ball{k5}
\Sigma_{ij}= \sigma \delta_{ij}+ \sigma' \hat b^i\hat b^j\,.
\gal
These terms describe the current flowing in the direction of the electric field and in the direction of the $\theta$-field gradient $\b b$:
\ball{k6}
\b j = \sigma \bm{\mathcal{E}}+ \sigma'(\unit b \cdot \bm{\mathcal{E}})\unit b\,.
\gal
The  two conductivities read
\bal
&\sigma= \frac{1}{2}(\Sigma\indices{^k_k}-\Sigma^{k\ell}\hat b_k\hat b_\ell)= \frac{e^2}{2T}\int f_0\nu^{-1} [v^2 -(\b v\cdot \unit b)^2] d^3p\,,\label{k7}\\
&\sigma'= \frac{1}{2}(3\Sigma^{k\ell}\hat b_k\hat b_\ell-\Sigma\indices{^k_k})= \frac{e^2}{2T}\int f_0\nu^{-1} [3(\b v\cdot \unit b)^2-v^2] d^3p\,.\label{k8}
\gal
In the case $\b b=0$, the integrands of \eq{k7} and \eq{k8} are spherically symmetric so that $\aver{v_iv_j}=v^2\delta_{ij}/3$, $\sigma'=0$ and $\sigma$ is given by \eq{k21}.

\begin{figure}[ht]
\begin{tabular}{cc}
      \includegraphics[height=5cm]{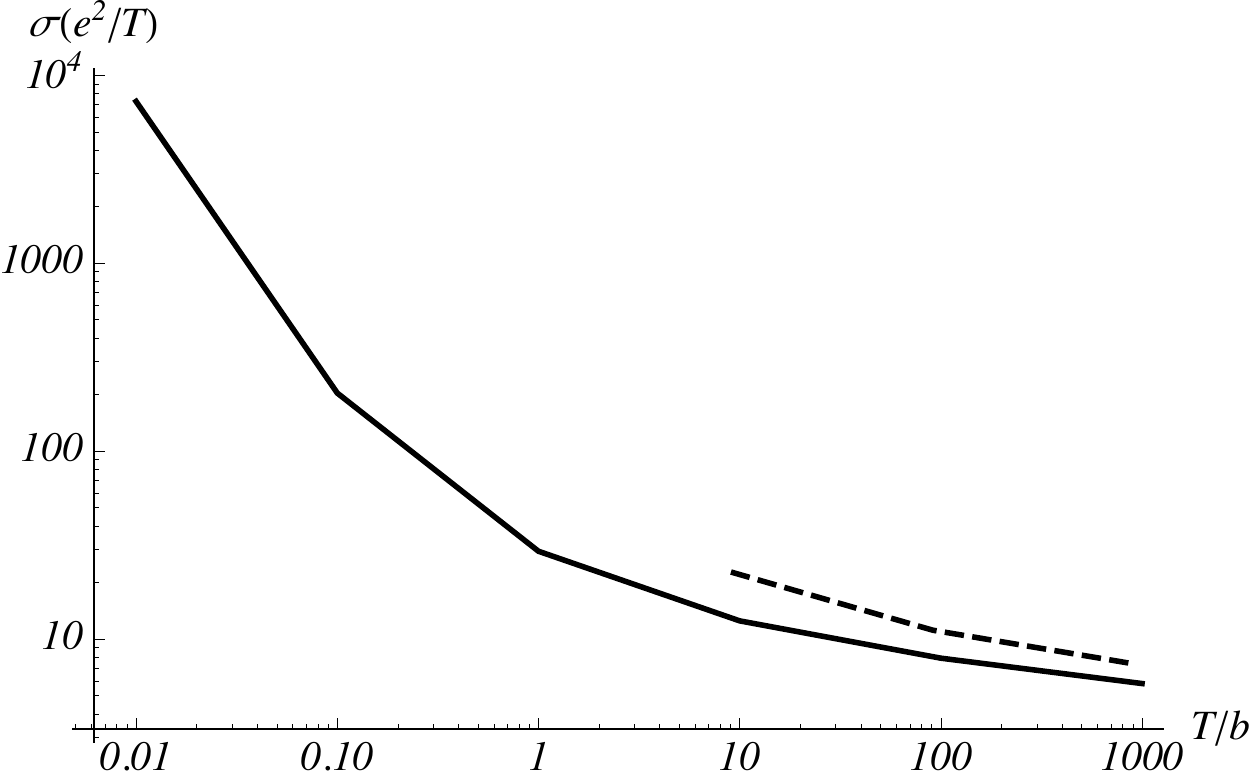} &
      \includegraphics[height=5cm]{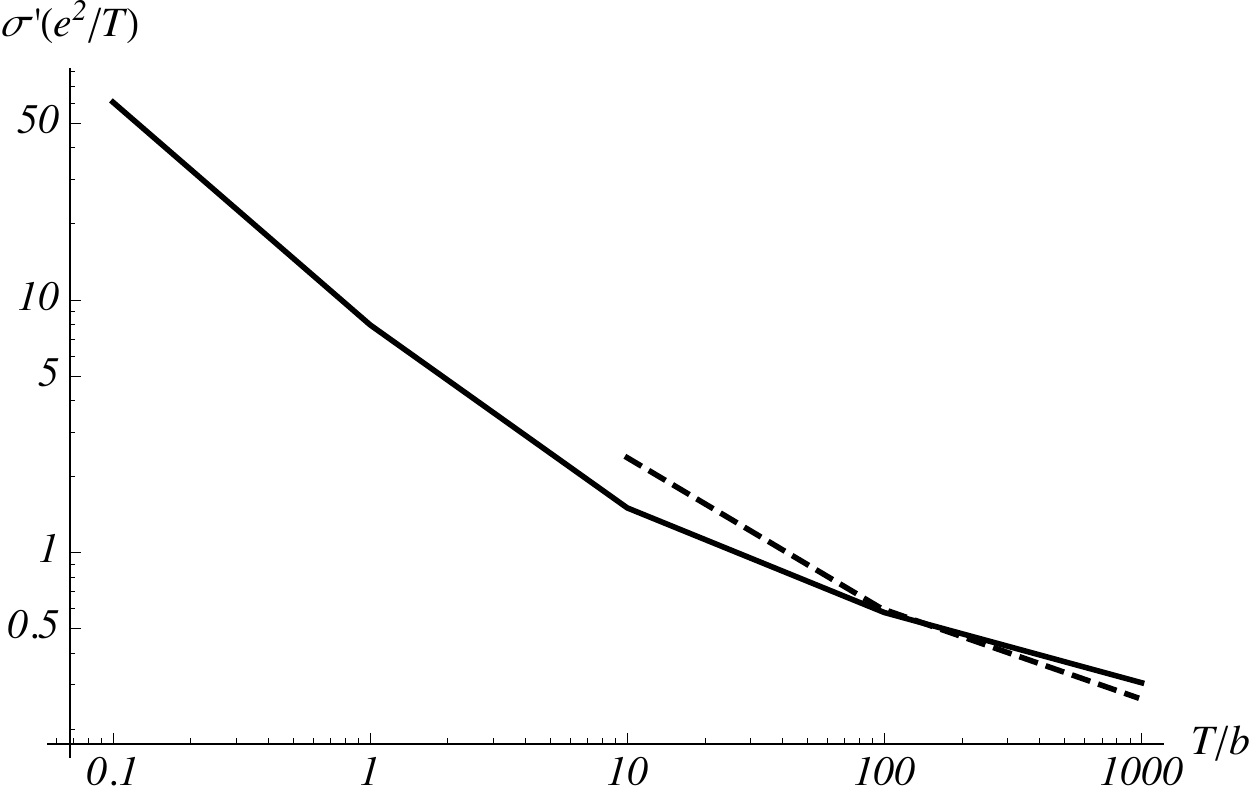}
      \end{tabular}
  \caption{Electrical conductivities \eq{k7} and \eq{k8} (solid lines) and their high temperature approximations \eq{k11} (dashed lines).  }
\label{fig5}
\end{figure}

In the limit $T\gg |\b b|$, the main contribution to these integrals arises from $|\b p|\gg |\b b|$ region where the integrals can be evaluated analytically. Substituting \eq{c10} into \eq{k7} and \eq{k8} we get in the chiral limit 
\ball{k11} 
\sigma= \frac{16\pi T}{e^2\ln (T/|\b b|)}\,,\qquad \sigma'= \frac{4\pi T}{e^2\ln^2 (T/|\b b|)}\,.
\gal
We observe the logarithmic suppression of electric current in the $\b b$ direction as compared to the electric field direction. The temperature dependence of the conductivities is displayed in \fig{fig5}. The suppression effect that the resonance at finite $b_0$ has on the magnitude of the electric conductivity can be appreciated by comparing the left panel of \fig{fig5} and \fig{fig9} at $T\gg b_0$ and $T\gg |\b b|$ respectively. 

\section{Discussion and Conclusions}\label{sec:s}

The main feature of the electron-ion scattering differential cross section at finite $b_0$ is the resonance, which appears as the pole of the photon propagator \eq{b10} and is intimately related to the chiral instability of the electromagnetic field. It exists  at any collision energy but its particular form depends of course on the collision kinematics. In the case of electron scattering on a heavy ion,  the resonance scattering angle is found to be  $\vartheta=b_0/|\b p|$ assuming that $\b b=0$. The same result also holds in the ultrarelativistic limit when $E\approx E'$ (or, more precisely, $|q^0|\ll 2|\b p|\sin\frac{\vartheta}{2}$). 
At finite $\b b$ the denominator of \eq{b10}  is proportional to $\b q^2+\b b^2\sin^2\alpha-b_0^2$ where $\alpha$ is the angle between $\b b$ and $\b q$. Clearly, the resonance emerges at any finite $b_0$ regardless of the magnitude of $\b b$, though at large $\b b$ it is confined to small values of $\sin\alpha$. The magnitude of the cross section at the resonance depends on the mechanism that tames the chiral instability of the electromagnetic field. Nevertheless, the transport cross section at high momenta given by \eq{d12}, is not sensitive to it  and is only logarithmically sensitive to $b_0$.

It was pointed out in \cite{Khaidukov:2013sja,Qiu:2016hzd} that the third term in \eq{b12c} is topological in the sense that the corresponding contribution to the interaction energy of two current loops is proportional to their Gauss linking number. The topological term is certainly present in the scattering amplitude describing interaction of the electron current with the ion magnetic moment. However, once the polarization sum is performed to obtain \eq{c3} the topological contribution can not be easily identified and is, actually, not specially interesting in the scattering problem. Nevertheless, it is of interest in bound states, which can also be addressed using similar methods. 

In conclusion, we computed the scattering and transport cross sections in chiral matter. The scattering is resonant at the scattering angles $\vartheta\approx b_0/|\b p|$, which implies suppression of the transport coefficients at temperatures higher than the chiral conductivity $T>b_0$.  At finite $\b b$ the transport is anisotropic as it acquires a component in the $\b b$-direction.  These observations provide a new avenue for experimental investigation of the chiral anomaly in chiral matter. In particular, their generalization to QCD may explain  the unusually small values of the shear viscosity to the entropy density ratio of the Quark-Gluon Plasma.

\acknowledgments
This work  was supported in part by the U.S. Department of Energy under Grant No.\ DE-FG02-87ER40371.

\appendix
\section{Fermion dispersion relation}\label{sec:app}

To compute the electron dispersion relation at finite $b^\mu$ we start from the Lagrangian describing the coupling of the axial current to the $\theta$-field
\ball{j1}
\mathcal{L}= \bar \psi (i\slashed{\partial} - \slashed{b}\gamma^5-m)\psi\,.
\gal
The corresponding equation of motion is
\ball{j2}
(i\slashed{\partial} - \slashed{b}\gamma^5-m)\psi=0\,.
\gal
Acting with the operator $(i\slashed{\partial}  - \slashed{b}\gamma^5+m)$ on  \eq{j2} yields
\ball{j4}
p^2-b^2-m^2+2\gamma^5\sigma^{\mu\nu}b_\mu p_\nu=0\,,
\gal
where $\sigma^{\mu\nu}= [\gamma^\mu,\gamma^\nu]/2$.  To get rid of the Dirac matrices, apply the operator $p^2-b^2-m^2-2\gamma^5\sigma^{\mu\nu}b_\mu p_\nu$ with the result
\ball{j5}
[(p^2-b^2-m^2)^2-2(\epsilon^{\mu\nu\lambda\rho}b_\mu p_\nu)^2]\psi=0\,.
\gal
The corresponding dispersion relation can written down as 
\ball{j8}
(p+b)^2(p-b)^2-2m^2(p^2-b^2)+m^4=0\,.
\gal
As before, we will consider two special cases. If  $b_0=0$ then the dispersion relation reads
\ball{j10}
E^2= \b p^2+\b b^2+m^2\pm 2\sqrt{m^2\b b^2+(\b p\cdot \b b)^2}\,,
\gal
where the two signs correspond to two helicity states. If $\b b=0$, then  the dispersion relation becomes
\ball{j11}
E^2= \b p^2+m^2-b_0^2\pm 2b_0\sqrt{E^2-m^2}\,.
\gal

In the chiral limit \eq{j8} reduces to  $(p\pm b)^2=0$ which implies
\ball{j13}
(E\pm b_0)^2=(\b p\mp \b b)^2\,.
\gal
In fact, these relations can be easily derived directly from \eq{j2} by writing it in the form $i\dot \psi = H\psi$ with the Hamiltonian 
\ball{j15}
H= \b \alpha\cdot (\b p+ \gamma^5\b b ) +b_0 \gamma^5 +m\gamma^0\,,
\gal 
and the usual notations for the Dirac matrices. Splitting the bispinor $\psi$ into two-component spinors $\psi= (u_L,u_R)e^{-ip\cdot x}$ we obtain in the chiral limit in the chiral representation 
\begin{subequations}\label{j17}
\bal
&[E-b_0-\b \sigma\cdot (\b p+\b b)]u_R=0\,,\label{j17a}\\
&[E+b_0+\b \sigma\cdot (\b p-\b b)]u_L=0\,,\label{j17b}
\gal
\end{subequations}
which yields \eq{j13}. 

Apparently, the wave function of a massless fermion at finite $b^\mu$ can be obtained from a free fermion wave function  by a substitution $E\to E_\pm = E\pm b_0$ and $\b p\to \b p_\mp = \b p\mp \b b$. In \sec{sec:b} and \sec{sec:d} we computed the cross sections for free fermions. Taking the $b$-contribution in the fermion wave function into account amounts to replacing $E\to E_\pm$ and $\b p\to \b p_\mp$ in  \eq{c8}, \eq{d1}, \eq{d5} etc. Thus, we are dealing with a theory of effectively free massless fermions of energy $E_\pm$ and momentum $p_\mp$ for right and left-handed states obeying the dispersion relation $E_\pm= |\b p_\mp|$.

The equilibrium distribution in the chiral limit reads
\ball{j20}
f_{0\pm}= \frac{n_\pm}{8\pi T^3}e^{-\beta E_\pm}\,,
\gal
where $n_\pm$ is the fermion number density. The phase space integrals are done over $\b p_\pm$. For example, the normalization condition of \eq{j20} is $\int f_{0\pm} d^3p_\pm =n_\pm$. Clearly, one can change the integration variable $\b p_\pm \to \b p$. Analysis of \sec{sec:j} assumes that the number densities of the two helicity states are the same. Thus, all indices $\pm$ can be omitted altogether.  

In summary, in order to compute the transport coefficients in the chiral limit, one can ignore the modification of the fermion wave functions at finite $b^\mu$ and consider distribution and scattering of free fermions.  The finite $m$ corrections may can be computed using \eq{j10} and \eq{j11}, but they, as well as the Berry curvature terms \cite{Son:2012zy}, are beyond the scope of this paper.


\end{document}